\documentstyle[12pt]{article}
\begin{document}
\begin{center}
{\huge\bf Damage spreading in the 'sandpile' model of SOC}
\end{center}
\vspace {0.4 cm}
\begin{center} {\large Ajanta Bhowal$^*$} \end{center}
\vspace {0.3 cm}
\begin{center} {\it Saha Institute of Nuclear Physics} \end{center}
\begin{center} {\it 1/AF Bidhannagar, Calcutta-700064, India} \end{center}
\vspace {1.0 cm}
\noindent {\bf Abstract:} We have studied the damage spreading (defined
in the text) in the 'sandpile' model of self organised criticality. We have
studied the variations of the critical time (defined in the text) and
the total number of sites damaged at critical time 
as a function of system size. Both shows the power law variation.

\vspace {1 cm}
\leftline {\bf Keywords: Sandpile model of SOC, Damage spreading}
\leftline {\bf PACS Numbers: 05.50 +q}

\vspace {3 cm}
\leftline {----------------------------}
\leftline {{\bf $^*$Present address:}Department of Physics, Lady Brabourne
College,}
\leftline {P-1/2 Suhrawardy Avenue, Calcutta-700017, India}
\leftline {E-mail:ajanta.bhowal@gmail.com}
\newpage

\leftline {\bf I. Introduction}

The damage spreading [1-3] studies the dynamic behaviour of
cooperative systems. The main idea of this
problem is to study how a small perturbation, called a damage,
in a cooperative system changes with the evolution of time.
This is studied  by observing the time evolution of the
two copies of the system  with slightly different initial 
configuration under the same dynamics and measuring the
damage by counting the number of elements which are different in the 
two copies of the system.
The damage spreading has been studied exhaustively,
in the  
Kauffman model \cite{ak}
and the spin systems [2,3].

In this paper, we have studied, by computer simulation, 
how damage spreads, in the 'sandpile' model,
over the whole lattice during the course of time evolution. 
'sandpile' model  \cite{btw} is a
lattice automata model which describes the appearance of 
long range spatio-temporal correlations observed in extended,
dissipative dynamical systems in nature. The essential
feature of this model is the occurence of fractal structure  in
space and '$1/f$' noise  in time, which is so called self-organised
criticality (SOC) \cite{btw}.  Substantial developments have
been made on the study of 'sandpile' model. 
But all these studies have been made in the steady (SOC) state,
reached by the system.

\bigskip
\leftline {\bf II. The model and simulation}

The lattice automata 'sandpile' model \cite{btw} 
of SOC evolves to a stationary
state in a self-organised (having no tunable parameter) way. This state has
no scale of length and time, hence is called critical. Altogether the state is
called self-organised critical state.  The description of the lattice automata 
model is as follows: At each site of this lattice, a variable
(automaton) $z(i,j)$ is associated which can take positive integer values.
Starting from the initial condition (at every site $z(i,j)$ = 0), the value 
of $z(i,j)$ is increased (so called addition of one 'sand' particle) at 
a randomly chosen site $(i,j)$ of the lattice in steps of unity as,
$$z(i,j) = z(i,j) + 1.$$
\noindent When the value of $z$ at any site reaches a maximum $z_m$, its
value decreases by four units (i.e., it topples) and each of the four nearest
neighbours gets one unit of $z$ (maintaining local conservation) as follows:
$$z(i,j) = z(i,j) - 4$$
$$z(i\pm 1, j\pm 1) = z(i\pm 1, j\pm 1) + 1 \eqno(1)$$
\noindent for $z(i,j) \geq z_m$. At the boundary sites $z$ = 0 (dissipative;
open boundary condition).

In this simulation, a square lattice of size $L\times L$ has been considered.
The value of $z_m$ = 4 here. It has been observed that, as the time goes on the 
average value ($\bar z$) of $z(i,j)$, over the space, increases and ultimately
reaches a steady value ($\bar z_c$) characterising the SOC state. 

We study here, by computer simulation, 
how a small perturbation spreads in time in the 'sandplie' model.
We have considered a square lattice and allowed it to evolve 
under the dynamics until the average value 
($\bar z$) of $z(i,j)$  reaches a steady value ($\bar z_c$).
We also have considered a 2nd lattice, 
which is a replica of the 1st lattice.
After reaching the steady (SOC) state (characterised by the steady value
of $\bar z$) we have perturbed suddenly one of the system (say the 1st lattice)
by adding unity to the automaton value at the central site of the lattice,
i.e,
$z_1(l/2,l/2)=z_1(l/2,l/2)+1$ .
Then we allowed the two lattices to evolve in time by the specified dynamics 
in the same way (i.e, by using the same sequence of random
numbers). It will be observed that both  
lattices (perturbed and unperturbed) give
the same macroscopic behavior (the same 
$\bar z_c$ and same scale invariant (power law) distribution of the 
avalanches size). But the microscopic details (the $z(i,j)$ at any
site $i,j$ at any time) of 
the two lattices are different. The 
differences in microscopic details of the two 
lattice are described here in terms of the 
"damaged sites", i.e., the sites of the perturbed 
lattice which are different from the 
unperturbed one. The damaged lattice is characterised 
by a variable (say $d(i,j)$),
which is zero  if the two lattices have the same $z$ 
(for any site $i,j$) value  and  1 otherwise. 
More precisely, $d(i,j) = 0$  if $z_1(i,j)=z_2(i,j)$ and $d(i,j)
= 1$ otherwise.  The non-zero sites of the damaged lattice (i.e., $d(i,j)=1$)
indicates the damage in this model.

\bigskip
\leftline {\bf III. Results}

It has been observed that,  with the evolution of time, 
the damage (cloud formed by the sites having $d(i,j) = 1$)
spreads and touches any one of the boundary line of the lattice
at time $\tau$ (starting from the initial time when the central
site of lattice 1 was perturbed by adding unity).
Here we study the  spreading of damage by measuring the following 
quantities:

\noindent (1) Minimum time ($\tau$) taken by the 
damage to touch any one of the boundaries (upper or lower)
of the lattice.

\noindent (2) The number ($M_{\tau}$) of damaged sites at $\tau$-th instant.

\noindent (3) The total number ($M_{Tot}$) of sites damaged during the
time $\tau$, starting from the initial time when the perturbation was
added.

Fig.1 shows the variation of critical time $(\tau )$ with the sytem size
$L$ in the log scale indicating the power law variation, following,
$\tau \sim L^a$ with $a = 1.97 $. 
Fig.2 shows the variations of $M_{\tau}$, the number of sites damaged at critical
time $(\tau )$ and  $M_{tot}$, the total number of sites damaged up to critical 
time with the syatem size $L$ in the log scale, 
indicating the power law
variations, $M_{\tau} \sim L^b$ with $b=1.47$.
and  $M_{tot} \sim L^c$ with $c=1.66$.
Thus we see that $\tau$, $M_{\tau}$ and $M_{tot}$, all these
quantities follow  a power law variation with the system size.
 All these data have been obtained by sampling over 100 different
seeds of random number generator for $L=30,40,\cdots 150$ and over
10 samples for $L=200,300$.

\bigskip
\leftline {\bf IV. Summary}

In the case of damage spreading in Ising model the spreading of damage is 
controlled by tuning the temperature. But due to the absence of any 
tunable parameter in the 'sandpile' model
it has been observed that in this case there is number spreading transition
 contrast to the other cases of damage spreading (like Ising system 
etc).  There is a  power law variation of critical time ($\tau$) 
and $M_{\tau}$, $M_{tot}$ with length as in the other cases (though the
exponents are different).

\newpage
\setlength{\unitlength}{0.240900pt}
\ifx\plotpoint\undefined\newsavebox{\plotpoint}\fi
\sbox{\plotpoint}{\rule[-0.200pt]{0.400pt}{0.400pt}}%
\begin{picture}(1500,900)(0,0)
\font\gnuplot=cmr10 at 10pt
\gnuplot
\sbox{\plotpoint}{\rule[-0.200pt]{0.400pt}{0.400pt}}%
\put(220.0,113.0){\rule[-0.200pt]{4.818pt}{0.400pt}}
\put(198,113){\makebox(0,0)[r]{1000}}
\put(1416.0,113.0){\rule[-0.200pt]{4.818pt}{0.400pt}}
\put(220.0,190.0){\rule[-0.200pt]{2.409pt}{0.400pt}}
\put(1426.0,190.0){\rule[-0.200pt]{2.409pt}{0.400pt}}
\put(220.0,235.0){\rule[-0.200pt]{2.409pt}{0.400pt}}
\put(1426.0,235.0){\rule[-0.200pt]{2.409pt}{0.400pt}}
\put(220.0,266.0){\rule[-0.200pt]{2.409pt}{0.400pt}}
\put(1426.0,266.0){\rule[-0.200pt]{2.409pt}{0.400pt}}
\put(220.0,291.0){\rule[-0.200pt]{2.409pt}{0.400pt}}
\put(1426.0,291.0){\rule[-0.200pt]{2.409pt}{0.400pt}}
\put(220.0,311.0){\rule[-0.200pt]{2.409pt}{0.400pt}}
\put(1426.0,311.0){\rule[-0.200pt]{2.409pt}{0.400pt}}
\put(220.0,328.0){\rule[-0.200pt]{2.409pt}{0.400pt}}
\put(1426.0,328.0){\rule[-0.200pt]{2.409pt}{0.400pt}}
\put(220.0,343.0){\rule[-0.200pt]{2.409pt}{0.400pt}}
\put(1426.0,343.0){\rule[-0.200pt]{2.409pt}{0.400pt}}
\put(220.0,356.0){\rule[-0.200pt]{2.409pt}{0.400pt}}
\put(1426.0,356.0){\rule[-0.200pt]{2.409pt}{0.400pt}}
\put(220.0,368.0){\rule[-0.200pt]{4.818pt}{0.400pt}}
\put(198,368){\makebox(0,0)[r]{10000}}
\put(1416.0,368.0){\rule[-0.200pt]{4.818pt}{0.400pt}}
\put(220.0,444.0){\rule[-0.200pt]{2.409pt}{0.400pt}}
\put(1426.0,444.0){\rule[-0.200pt]{2.409pt}{0.400pt}}
\put(220.0,489.0){\rule[-0.200pt]{2.409pt}{0.400pt}}
\put(1426.0,489.0){\rule[-0.200pt]{2.409pt}{0.400pt}}
\put(220.0,521.0){\rule[-0.200pt]{2.409pt}{0.400pt}}
\put(1426.0,521.0){\rule[-0.200pt]{2.409pt}{0.400pt}}
\put(220.0,546.0){\rule[-0.200pt]{2.409pt}{0.400pt}}
\put(1426.0,546.0){\rule[-0.200pt]{2.409pt}{0.400pt}}
\put(220.0,566.0){\rule[-0.200pt]{2.409pt}{0.400pt}}
\put(1426.0,566.0){\rule[-0.200pt]{2.409pt}{0.400pt}}
\put(220.0,583.0){\rule[-0.200pt]{2.409pt}{0.400pt}}
\put(1426.0,583.0){\rule[-0.200pt]{2.409pt}{0.400pt}}
\put(220.0,598.0){\rule[-0.200pt]{2.409pt}{0.400pt}}
\put(1426.0,598.0){\rule[-0.200pt]{2.409pt}{0.400pt}}
\put(220.0,611.0){\rule[-0.200pt]{2.409pt}{0.400pt}}
\put(1426.0,611.0){\rule[-0.200pt]{2.409pt}{0.400pt}}
\put(220.0,622.0){\rule[-0.200pt]{4.818pt}{0.400pt}}
\put(198,622){\makebox(0,0)[r]{100000}}
\put(1416.0,622.0){\rule[-0.200pt]{4.818pt}{0.400pt}}
\put(220.0,699.0){\rule[-0.200pt]{2.409pt}{0.400pt}}
\put(1426.0,699.0){\rule[-0.200pt]{2.409pt}{0.400pt}}
\put(220.0,744.0){\rule[-0.200pt]{2.409pt}{0.400pt}}
\put(1426.0,744.0){\rule[-0.200pt]{2.409pt}{0.400pt}}
\put(220.0,776.0){\rule[-0.200pt]{2.409pt}{0.400pt}}
\put(1426.0,776.0){\rule[-0.200pt]{2.409pt}{0.400pt}}
\put(220.0,800.0){\rule[-0.200pt]{2.409pt}{0.400pt}}
\put(1426.0,800.0){\rule[-0.200pt]{2.409pt}{0.400pt}}
\put(220.0,821.0){\rule[-0.200pt]{2.409pt}{0.400pt}}
\put(1426.0,821.0){\rule[-0.200pt]{2.409pt}{0.400pt}}
\put(220.0,838.0){\rule[-0.200pt]{2.409pt}{0.400pt}}
\put(1426.0,838.0){\rule[-0.200pt]{2.409pt}{0.400pt}}
\put(220.0,852.0){\rule[-0.200pt]{2.409pt}{0.400pt}}
\put(1426.0,852.0){\rule[-0.200pt]{2.409pt}{0.400pt}}
\put(220.0,865.0){\rule[-0.200pt]{2.409pt}{0.400pt}}
\put(1426.0,865.0){\rule[-0.200pt]{2.409pt}{0.400pt}}
\put(220.0,877.0){\rule[-0.200pt]{4.818pt}{0.400pt}}
\put(198,877){\makebox(0,0)[r]{1e+06}}
\put(1416.0,877.0){\rule[-0.200pt]{4.818pt}{0.400pt}}
\put(220.0,113.0){\rule[-0.200pt]{0.400pt}{4.818pt}}
\put(220,68){\makebox(0,0){10}}
\put(220.0,857.0){\rule[-0.200pt]{0.400pt}{4.818pt}}
\put(403.0,113.0){\rule[-0.200pt]{0.400pt}{2.409pt}}
\put(403.0,867.0){\rule[-0.200pt]{0.400pt}{2.409pt}}
\put(510.0,113.0){\rule[-0.200pt]{0.400pt}{2.409pt}}
\put(510.0,867.0){\rule[-0.200pt]{0.400pt}{2.409pt}}
\put(586.0,113.0){\rule[-0.200pt]{0.400pt}{2.409pt}}
\put(586.0,867.0){\rule[-0.200pt]{0.400pt}{2.409pt}}
\put(645.0,113.0){\rule[-0.200pt]{0.400pt}{2.409pt}}
\put(645.0,867.0){\rule[-0.200pt]{0.400pt}{2.409pt}}
\put(693.0,113.0){\rule[-0.200pt]{0.400pt}{2.409pt}}
\put(693.0,867.0){\rule[-0.200pt]{0.400pt}{2.409pt}}
\put(734.0,113.0){\rule[-0.200pt]{0.400pt}{2.409pt}}
\put(734.0,867.0){\rule[-0.200pt]{0.400pt}{2.409pt}}
\put(769.0,113.0){\rule[-0.200pt]{0.400pt}{2.409pt}}
\put(769.0,867.0){\rule[-0.200pt]{0.400pt}{2.409pt}}
\put(800.0,113.0){\rule[-0.200pt]{0.400pt}{2.409pt}}
\put(800.0,867.0){\rule[-0.200pt]{0.400pt}{2.409pt}}
\put(828.0,113.0){\rule[-0.200pt]{0.400pt}{4.818pt}}
\put(828,68){\makebox(0,0){100}}
\put(828.0,857.0){\rule[-0.200pt]{0.400pt}{4.818pt}}
\put(1011.0,113.0){\rule[-0.200pt]{0.400pt}{2.409pt}}
\put(1011.0,867.0){\rule[-0.200pt]{0.400pt}{2.409pt}}
\put(1118.0,113.0){\rule[-0.200pt]{0.400pt}{2.409pt}}
\put(1118.0,867.0){\rule[-0.200pt]{0.400pt}{2.409pt}}
\put(1194.0,113.0){\rule[-0.200pt]{0.400pt}{2.409pt}}
\put(1194.0,867.0){\rule[-0.200pt]{0.400pt}{2.409pt}}
\put(1253.0,113.0){\rule[-0.200pt]{0.400pt}{2.409pt}}
\put(1253.0,867.0){\rule[-0.200pt]{0.400pt}{2.409pt}}
\put(1301.0,113.0){\rule[-0.200pt]{0.400pt}{2.409pt}}
\put(1301.0,867.0){\rule[-0.200pt]{0.400pt}{2.409pt}}
\put(1342.0,113.0){\rule[-0.200pt]{0.400pt}{2.409pt}}
\put(1342.0,867.0){\rule[-0.200pt]{0.400pt}{2.409pt}}
\put(1377.0,113.0){\rule[-0.200pt]{0.400pt}{2.409pt}}
\put(1377.0,867.0){\rule[-0.200pt]{0.400pt}{2.409pt}}
\put(1408.0,113.0){\rule[-0.200pt]{0.400pt}{2.409pt}}
\put(1408.0,867.0){\rule[-0.200pt]{0.400pt}{2.409pt}}
\put(1436.0,113.0){\rule[-0.200pt]{0.400pt}{4.818pt}}
\put(1436,68){\makebox(0,0){1000}}
\put(1436.0,857.0){\rule[-0.200pt]{0.400pt}{4.818pt}}
\put(220.0,113.0){\rule[-0.200pt]{292.934pt}{0.400pt}}
\put(1436.0,113.0){\rule[-0.200pt]{0.400pt}{184.048pt}}
\put(220.0,877.0){\rule[-0.200pt]{292.934pt}{0.400pt}}
\put(45,495){\makebox(0,0){$\tau$}}
\put(828,23){\makebox(0,0){L}}
\put(220.0,113.0){\rule[-0.200pt]{0.400pt}{184.048pt}}
\put(510,201){\raisebox{-.8pt}{\makebox(0,0){$\Diamond$}}}
\put(586,264){\raisebox{-.8pt}{\makebox(0,0){$\Diamond$}}}
\put(645,313){\raisebox{-.8pt}{\makebox(0,0){$\Diamond$}}}
\put(693,351){\raisebox{-.8pt}{\makebox(0,0){$\Diamond$}}}
\put(734,386){\raisebox{-.8pt}{\makebox(0,0){$\Diamond$}}}
\put(769,414){\raisebox{-.8pt}{\makebox(0,0){$\Diamond$}}}
\put(800,438){\raisebox{-.8pt}{\makebox(0,0){$\Diamond$}}}
\put(828,461){\raisebox{-.8pt}{\makebox(0,0){$\Diamond$}}}
\put(935,555){\raisebox{-.8pt}{\makebox(0,0){$\Diamond$}}}
\put(1011,616){\raisebox{-.8pt}{\makebox(0,0){$\Diamond$}}}
\put(1118,701){\raisebox{-.8pt}{\makebox(0,0){$\Diamond$}}}
\sbox{\plotpoint}{\rule[-0.400pt]{0.800pt}{0.800pt}}%
\put(510,201){\usebox{\plotpoint}}
\multiput(510.00,202.41)(0.606,0.500){997}{\rule{1.169pt}{0.120pt}}
\multiput(510.00,199.34)(605.574,502.000){2}{\rule{0.584pt}{0.800pt}}
\end{picture}
\vskip 0.5 cm
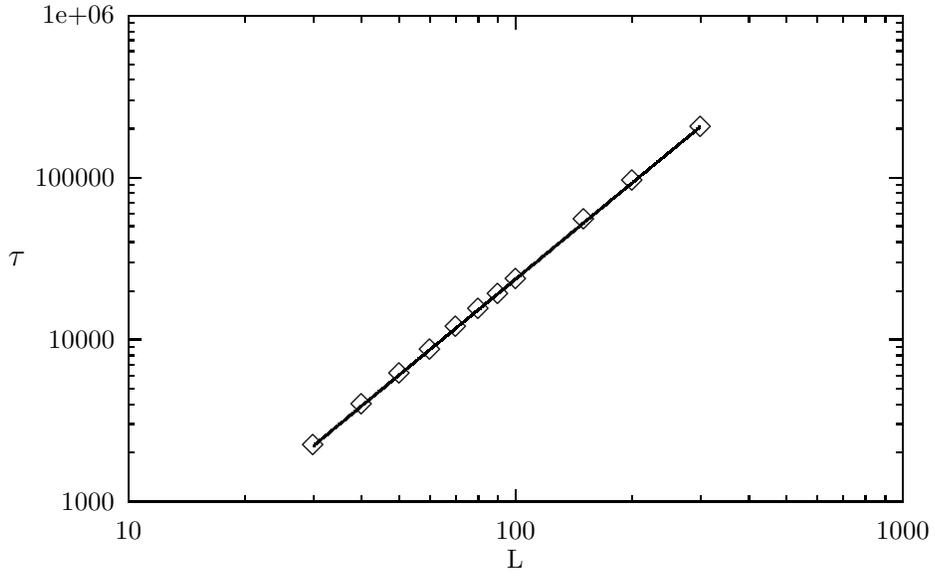
\leftline {{\bf Fig. 1.} Variation of critical time $(\tau )$  with
system size $L$.}
\vskip 1.2 cm
\setlength{\unitlength}{0.240900pt}
\ifx\plotpoint\undefined\newsavebox{\plotpoint}\fi
\sbox{\plotpoint}{\rule[-0.200pt]{0.400pt}{0.400pt}}%
\begin{picture}(1500,900)(0,0)
\font\gnuplot=cmr10 at 10pt
\gnuplot
\sbox{\plotpoint}{\rule[-0.200pt]{0.400pt}{0.400pt}}%
\put(176.0,113.0){\rule[-0.200pt]{4.818pt}{0.400pt}}
\put(154,113){\makebox(0,0)[r]{100}}
\put(1416.0,113.0){\rule[-0.200pt]{4.818pt}{0.400pt}}
\put(176.0,190.0){\rule[-0.200pt]{2.409pt}{0.400pt}}
\put(1426.0,190.0){\rule[-0.200pt]{2.409pt}{0.400pt}}
\put(176.0,235.0){\rule[-0.200pt]{2.409pt}{0.400pt}}
\put(1426.0,235.0){\rule[-0.200pt]{2.409pt}{0.400pt}}
\put(176.0,266.0){\rule[-0.200pt]{2.409pt}{0.400pt}}
\put(1426.0,266.0){\rule[-0.200pt]{2.409pt}{0.400pt}}
\put(176.0,291.0){\rule[-0.200pt]{2.409pt}{0.400pt}}
\put(1426.0,291.0){\rule[-0.200pt]{2.409pt}{0.400pt}}
\put(176.0,311.0){\rule[-0.200pt]{2.409pt}{0.400pt}}
\put(1426.0,311.0){\rule[-0.200pt]{2.409pt}{0.400pt}}
\put(176.0,328.0){\rule[-0.200pt]{2.409pt}{0.400pt}}
\put(1426.0,328.0){\rule[-0.200pt]{2.409pt}{0.400pt}}
\put(176.0,343.0){\rule[-0.200pt]{2.409pt}{0.400pt}}
\put(1426.0,343.0){\rule[-0.200pt]{2.409pt}{0.400pt}}
\put(176.0,356.0){\rule[-0.200pt]{2.409pt}{0.400pt}}
\put(1426.0,356.0){\rule[-0.200pt]{2.409pt}{0.400pt}}
\put(176.0,368.0){\rule[-0.200pt]{4.818pt}{0.400pt}}
\put(154,368){\makebox(0,0)[r]{1000}}
\put(1416.0,368.0){\rule[-0.200pt]{4.818pt}{0.400pt}}
\put(176.0,444.0){\rule[-0.200pt]{2.409pt}{0.400pt}}
\put(1426.0,444.0){\rule[-0.200pt]{2.409pt}{0.400pt}}
\put(176.0,489.0){\rule[-0.200pt]{2.409pt}{0.400pt}}
\put(1426.0,489.0){\rule[-0.200pt]{2.409pt}{0.400pt}}
\put(176.0,521.0){\rule[-0.200pt]{2.409pt}{0.400pt}}
\put(1426.0,521.0){\rule[-0.200pt]{2.409pt}{0.400pt}}
\put(176.0,546.0){\rule[-0.200pt]{2.409pt}{0.400pt}}
\put(1426.0,546.0){\rule[-0.200pt]{2.409pt}{0.400pt}}
\put(176.0,566.0){\rule[-0.200pt]{2.409pt}{0.400pt}}
\put(1426.0,566.0){\rule[-0.200pt]{2.409pt}{0.400pt}}
\put(176.0,583.0){\rule[-0.200pt]{2.409pt}{0.400pt}}
\put(1426.0,583.0){\rule[-0.200pt]{2.409pt}{0.400pt}}
\put(176.0,598.0){\rule[-0.200pt]{2.409pt}{0.400pt}}
\put(1426.0,598.0){\rule[-0.200pt]{2.409pt}{0.400pt}}
\put(176.0,611.0){\rule[-0.200pt]{2.409pt}{0.400pt}}
\put(1426.0,611.0){\rule[-0.200pt]{2.409pt}{0.400pt}}
\put(176.0,622.0){\rule[-0.200pt]{4.818pt}{0.400pt}}
\put(154,622){\makebox(0,0)[r]{10000}}
\put(1416.0,622.0){\rule[-0.200pt]{4.818pt}{0.400pt}}
\put(176.0,699.0){\rule[-0.200pt]{2.409pt}{0.400pt}}
\put(1426.0,699.0){\rule[-0.200pt]{2.409pt}{0.400pt}}
\put(176.0,744.0){\rule[-0.200pt]{2.409pt}{0.400pt}}
\put(1426.0,744.0){\rule[-0.200pt]{2.409pt}{0.400pt}}
\put(176.0,776.0){\rule[-0.200pt]{2.409pt}{0.400pt}}
\put(1426.0,776.0){\rule[-0.200pt]{2.409pt}{0.400pt}}
\put(176.0,800.0){\rule[-0.200pt]{2.409pt}{0.400pt}}
\put(1426.0,800.0){\rule[-0.200pt]{2.409pt}{0.400pt}}
\put(176.0,821.0){\rule[-0.200pt]{2.409pt}{0.400pt}}
\put(1426.0,821.0){\rule[-0.200pt]{2.409pt}{0.400pt}}
\put(176.0,838.0){\rule[-0.200pt]{2.409pt}{0.400pt}}
\put(1426.0,838.0){\rule[-0.200pt]{2.409pt}{0.400pt}}
\put(176.0,852.0){\rule[-0.200pt]{2.409pt}{0.400pt}}
\put(1426.0,852.0){\rule[-0.200pt]{2.409pt}{0.400pt}}
\put(176.0,865.0){\rule[-0.200pt]{2.409pt}{0.400pt}}
\put(1426.0,865.0){\rule[-0.200pt]{2.409pt}{0.400pt}}
\put(176.0,877.0){\rule[-0.200pt]{4.818pt}{0.400pt}}
\put(154,877){\makebox(0,0)[r]{100000}}
\put(1416.0,877.0){\rule[-0.200pt]{4.818pt}{0.400pt}}
\put(176.0,113.0){\rule[-0.200pt]{0.400pt}{4.818pt}}
\put(176,68){\makebox(0,0){10}}
\put(176.0,857.0){\rule[-0.200pt]{0.400pt}{4.818pt}}
\put(366.0,113.0){\rule[-0.200pt]{0.400pt}{2.409pt}}
\put(366.0,867.0){\rule[-0.200pt]{0.400pt}{2.409pt}}
\put(477.0,113.0){\rule[-0.200pt]{0.400pt}{2.409pt}}
\put(477.0,867.0){\rule[-0.200pt]{0.400pt}{2.409pt}}
\put(555.0,113.0){\rule[-0.200pt]{0.400pt}{2.409pt}}
\put(555.0,867.0){\rule[-0.200pt]{0.400pt}{2.409pt}}
\put(616.0,113.0){\rule[-0.200pt]{0.400pt}{2.409pt}}
\put(616.0,867.0){\rule[-0.200pt]{0.400pt}{2.409pt}}
\put(666.0,113.0){\rule[-0.200pt]{0.400pt}{2.409pt}}
\put(666.0,867.0){\rule[-0.200pt]{0.400pt}{2.409pt}}
\put(708.0,113.0){\rule[-0.200pt]{0.400pt}{2.409pt}}
\put(708.0,867.0){\rule[-0.200pt]{0.400pt}{2.409pt}}
\put(745.0,113.0){\rule[-0.200pt]{0.400pt}{2.409pt}}
\put(745.0,867.0){\rule[-0.200pt]{0.400pt}{2.409pt}}
\put(777.0,113.0){\rule[-0.200pt]{0.400pt}{2.409pt}}
\put(777.0,867.0){\rule[-0.200pt]{0.400pt}{2.409pt}}
\put(806.0,113.0){\rule[-0.200pt]{0.400pt}{4.818pt}}
\put(806,68){\makebox(0,0){100}}
\put(806.0,857.0){\rule[-0.200pt]{0.400pt}{4.818pt}}
\put(996.0,113.0){\rule[-0.200pt]{0.400pt}{2.409pt}}
\put(996.0,867.0){\rule[-0.200pt]{0.400pt}{2.409pt}}
\put(1107.0,113.0){\rule[-0.200pt]{0.400pt}{2.409pt}}
\put(1107.0,867.0){\rule[-0.200pt]{0.400pt}{2.409pt}}
\put(1185.0,113.0){\rule[-0.200pt]{0.400pt}{2.409pt}}
\put(1185.0,867.0){\rule[-0.200pt]{0.400pt}{2.409pt}}
\put(1246.0,113.0){\rule[-0.200pt]{0.400pt}{2.409pt}}
\put(1246.0,867.0){\rule[-0.200pt]{0.400pt}{2.409pt}}
\put(1296.0,113.0){\rule[-0.200pt]{0.400pt}{2.409pt}}
\put(1296.0,867.0){\rule[-0.200pt]{0.400pt}{2.409pt}}
\put(1338.0,113.0){\rule[-0.200pt]{0.400pt}{2.409pt}}
\put(1338.0,867.0){\rule[-0.200pt]{0.400pt}{2.409pt}}
\put(1375.0,113.0){\rule[-0.200pt]{0.400pt}{2.409pt}}
\put(1375.0,867.0){\rule[-0.200pt]{0.400pt}{2.409pt}}
\put(1407.0,113.0){\rule[-0.200pt]{0.400pt}{2.409pt}}
\put(1407.0,867.0){\rule[-0.200pt]{0.400pt}{2.409pt}}
\put(1436.0,113.0){\rule[-0.200pt]{0.400pt}{4.818pt}}
\put(1436,68){\makebox(0,0){1000}}
\put(1436.0,857.0){\rule[-0.200pt]{0.400pt}{4.818pt}}
\put(176.0,113.0){\rule[-0.200pt]{303.534pt}{0.400pt}}
\put(1436.0,113.0){\rule[-0.200pt]{0.400pt}{184.048pt}}
\put(176.0,877.0){\rule[-0.200pt]{303.534pt}{0.400pt}}
\put(806,23){\makebox(0,0){L}}
\put(176.0,113.0){\rule[-0.200pt]{0.400pt}{184.048pt}}
\put(477,211){\raisebox{-.8pt}{\makebox(0,0){$\Diamond$}}}
\put(555,251){\raisebox{-.8pt}{\makebox(0,0){$\Diamond$}}}
\put(616,292){\raisebox{-.8pt}{\makebox(0,0){$\Diamond$}}}
\put(666,309){\raisebox{-.8pt}{\makebox(0,0){$\Diamond$}}}
\put(708,345){\raisebox{-.8pt}{\makebox(0,0){$\Diamond$}}}
\put(745,363){\raisebox{-.8pt}{\makebox(0,0){$\Diamond$}}}
\put(777,393){\raisebox{-.8pt}{\makebox(0,0){$\Diamond$}}}
\put(806,398){\raisebox{-.8pt}{\makebox(0,0){$\Diamond$}}}
\put(917,477){\raisebox{-.8pt}{\makebox(0,0){$\Diamond$}}}
\put(996,529){\raisebox{-.8pt}{\makebox(0,0){$\Diamond$}}}
\put(1107,569){\raisebox{-.8pt}{\makebox(0,0){$\Diamond$}}}
\sbox{\plotpoint}{\rule[-0.400pt]{0.800pt}{0.800pt}}%
\put(477,207){\usebox{\plotpoint}}
\multiput(477.00,208.41)(0.840,0.500){743}{\rule{1.544pt}{0.121pt}}
\multiput(477.00,205.34)(626.795,375.000){2}{\rule{0.772pt}{0.800pt}}
\put(477,265){\makebox(0,0){$+$}}
\put(555,328){\makebox(0,0){$+$}}
\put(616,367){\makebox(0,0){$+$}}
\put(666,391){\makebox(0,0){$+$}}
\put(708,430){\makebox(0,0){$+$}}
\put(745,443){\makebox(0,0){$+$}}
\put(777,473){\makebox(0,0){$+$}}
\put(806,483){\makebox(0,0){$+$}}
\put(917,560){\makebox(0,0){$+$}}
\put(996,625){\makebox(0,0){$+$}}
\put(1107,692){\makebox(0,0){$+$}}
\put(477,269){\usebox{\plotpoint}}
\multiput(477.00,270.41)(0.745,0.500){839}{\rule{1.391pt}{0.121pt}}
\multiput(477.00,267.34)(627.112,423.000){2}{\rule{0.696pt}{0.800pt}}
\end{picture}
\vskip 0.5 cm
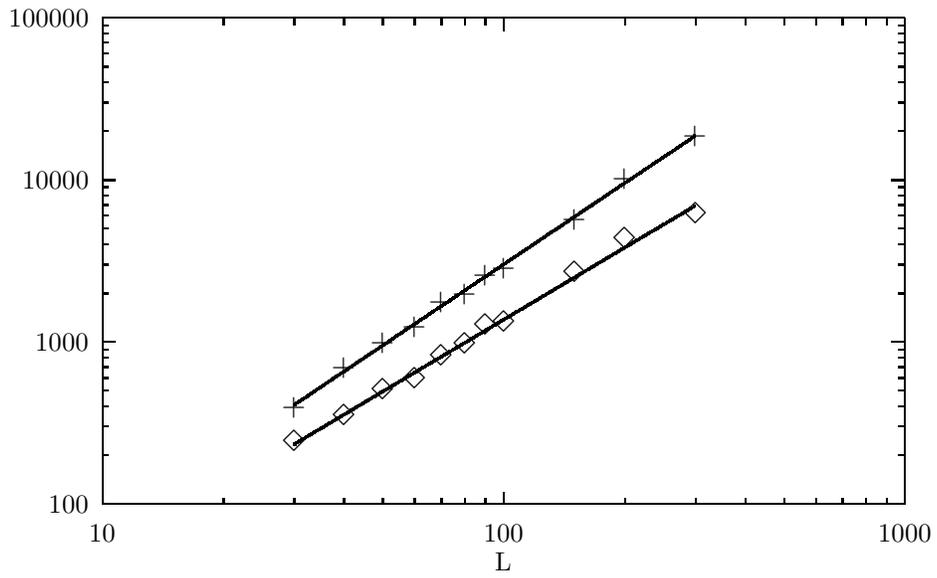
\leftline {{\bf Fig. 2.} Variations of $M_{\tau} 
 (\Diamond)$ and $M_{tot} (+)$
with system size $L$.}
\end{document}